\documentclass[twoside,openright,a4paper]{article}	
\usepackage{graphicx}	
\usepackage{graphics}	
\usepackage{epsfig}	
\usepackage{amsmath}	
\usepackage{amsfonts}	
\usepackage{sectsty}	
\sectionfont{\large}	
.360pk	
\font\rt=cmss9.360pk	
\font\sd=cmcsc9.360pk	

\makeatletter	
\renewcommand*\env@cases[1][1.5]
{
  \let\@ifnextchar\new@ifnextchar
  \left\lbrace
  \def\arraystretch{#1}
  \array{@{}l@{\quad}l@{}}
}
\makeatother

\begin{document}

\centerline{\Large Simple free-surface detection in two and three-dimensional SPH solver}

\medskip

\begin{center}
{\sc Agra Barecasco$^{1,2}$, Hanifa Terissa $^{1,2}$, Christian Fredy Naa$^2$}

\medskip
$^1$Graduate School of Natural Science and Technology, Kanazawa University\\ Kakuma, Kanazawa 920-1192 Japan
\\$^2$Faculty of Mathematics and Natural Sciences, Institut Teknologi Bandung\\ Jl. Ganesha 10, Bandung 40132 Indonesia
\\E-mail: agra.barecasco@students.itb.ac.id, hanifa.t@students.itb.ac.id, chris@cphys.fi.itb.ac.id
\end{center}

\bigskip

{\parindent 0pt
\textbf{Abstract.} {\em A simple free-surface particle detection method for two and three-dimensional SPH simulation has been implemented. The method uses sphere representation for the SPH particle. The fluid domain is covered by overlapping spheres. A sphere whose surface is not fully covered considered as boundary. To test particle's boundary status, we used a sum of normalized relative position vectors from neighbouring particles to the test particle. By checking the existence of uncovered sphere surface by this vector sum, boundary status of the test particle can be determined. This boundary detection method can be easily embedded in the SPH solver algorithm. }}

\section{Introduction}
	In recent years there has been much development of smoothed-particle hydrodynamics (SPH)\cite{Libersky}. Implementation of boundary conditions is not so clear with SPH method compared with mesh based methods. In SPH, the problem that we must first locate the points that are regarded as boundary. 
	
	For boundary particle detection, Randles and Libersky\cite{Libersky} previously have suggested using the sums of the gradients of the SPH kernels. Ideally, these kernel gradients sum to zero for interior particles. Any particle for which the sum of the gradient kernels is not near zero is presumably an exterior particle. This method gives correct results when the particles are uniformly spaced. Aamer and Dilts\cite{Dilts} developed an algorithm using overlapping spheres as fluid particles representation. The method can geometrically detect free-surface particles in a robust way. However, its extension to three dimensional simulations requires a large amount of calculation. Marrone\cite{Marrone} gave the idea of using a \textit{scan cone} around the expected normal vector of the fluid surface to make a further check if there is any particle covering the test particle.
	
	In this paper, we propose a simple boundary detection method using both the idea of Dilts\cite{Dilts} and Marrone\cite{Marrone}. The aim is to detect uncovered sphere segment on the boundary particle sphere. In our method, we develop a very simple algorithm to estimate the existence of this uncovered sphere segment.
	
\section{Numerical Schemes: Smoothed Particle Hydrodynamics}
	Smoothed-particle hydrodynamics is a method to simulate fluids\cite{Libersky}. It is a lagrangian meshless method based on convolution of smoothing function (\textit{kernel}) $W$ over fluid's field functions to approximate their values at a point. This field function is discretized by a set of point particles. The kernel approximation is given by the form
	
	\begin{equation*}
	f(\mathbf{x})=\int\limits_{\Omega}^{} f(\mathbf{x}')W(\mathbf{x}-\mathbf{x}')d\mathbf{x}',
	\end{equation*}
	
	where $\Omega$ represents the convolution domain. In our simulation we use the cubic spline kernel:
	\begin{equation*} 
	W(\mathbf{x}, h) = \beta
	\begin{cases}
		\frac{2}{3} - q^{2} + \frac{1}{2} q^{3}, &
			0 \le q < 1 \\
		\frac{1}{6}\left( 2 - q^{3} \right), &
			1 \le q < 2\\
		0, &
			q \ge 2,
	\end{cases}
	\end{equation*}
	where $2h$ is the radius of the kernel function's \textit{support} and $q = \frac{|\mathbf{x}|}{h}$. 
	
	To simulate an inviscid fluid, the convolution of this kernel is applied to the following governing Euler equations for inviscid fluid motion:
	\begin{align}
	\frac{\mbox{D}\rho}{\mbox{D}t} &= -\rho\nabla\cdot\mathbf{v}\\
	\frac{\mbox{D}\mathbf{v}}{\mbox{D}t} &= -\frac{1}{\rho}\nabla p + \mathbf{g}
	\end{align}
	where $\mbox{D}/\mbox{D}t$ represent material derivative following an infinitesemal fluid element. $\rho, \mathbf{v}, p$ represent density, velocity and pressure, respectively. The external acceleration $\mathbf{g}$ is given to simulate gravity.
	
	For SPH simulation, the fluid domain is discretized by fluid particles. The discretized SPH approximation form of Euler equations above are given by
	\begin{eqnarray}\label{eq:continuityDiscretized}
	\frac{\mbox{D}\rho_i}{\mbox{D}t} &=& \sum\limits_{j}^{} m_j\left( \mathbf{v}_i - \mathbf{v}_j \right) \cdot \nabla W(\mathbf{x}_i - \mathbf{x}_j, h)\\
	\label{eq:momentumDiscretized}
	\frac{\mbox{D}\mathbf{v}_i}{\mbox{D}t} &=& \sum\limits_{j}^{} m_j \left( \frac{p_i + p_j}{\rho_i\rho_j} \right) \nabla W(\mathbf{x}_i - \mathbf{x}_j, h) + \mathbf{g}, 
	\end{eqnarray}
	where $i$ and $j$ denote particle indices and $m_j$ represent $j$-th particle mass.
	
	To determine the pressure, we used the state equation 
	\begin{equation} \label{eq:stateEquation}
	p_i = c^2 (\rho_i - \rho_0),
	\end{equation}
	where $c$ is the speed of sound constant and $\rho_0$ is the reference density. Here $\mathbf{x}$ represents particle position and $j$ sums to all particles within the kernel radius. In the implementation, the sums must check for all particles in the fluid domain if they are within the kernel. To make a faster calculation, we used a linked-list grid method\cite{Liu} to reduce the calculation amount per particle so $j$ only runs through the nearest neighbouring particles.
	
	\subsection*{SPH solver algorithm}
	Implementation of eq.~(\ref{eq:continuityDiscretized}) and eq.~(\ref{eq:momentumDiscretized}) involves many steps. Here we give the general steps of SPH solver that are related to our boundary detection method. For our cases, we used Leap-Frog integrator to advance in time. SPH simulation can be run as follows:
	
	\begin{enumerate}
	\item \textit{Initialization}. Set the initial condition of attributes of all SPH particles
	\item Do the followings for all fluid particles until the run-time limit has been reached:
		\begin{enumerate}
		\item Calculate the current pressure of all particles by using state equation eq.~(\ref{eq:stateEquation})
		\item For each particle, insert it into the linked-list grid\cite{Liu} and register its nearest neighbour particles
		\item For each particle, calculate its momentum change rate (eq.~(\ref{eq:momentumDiscretized})) by using the new calculated pressure, then calculate its new velocity and position
		\item For each particle, calculate its density change rate (eq.~(\ref{eq:continuityDiscretized})) by using new calculated velocity, then calculate its new density
		\end{enumerate}
	\end{enumerate}
	
	We see that SPH simulation involves the calculation of summation from all neighbouring particles to determine each particle's density, pressure and momentum. This is an advantage for our detection method which is explained in the next section.

\section{Boundary Particle Detection Method}
	\subsection{Boundary particle definition}
	We are given a set of spheres in $\mathbb{R}^3$, $S=\{ s_1, s_2, \dots, s_n \}$. Each sphere $s_i$ represents a SPH particle centered at the particle's position $\mathbf{x}_i$. Let $r_i$ be the sphere's radius.
	
	A point $\mathbf{x}\in\mathbb{R}^3$ is said to be \emph{covered} by $s_i$ if $|\mathbf{x}-\mathbf{x}_i|<r_i$. SPH \emph{fluid domain} $F_S\in\mathbb{R}^3$ is a domain where every $\mathbf{x}\in F_S$ is covered by one or more SPH spheres of $S$. $F_S$ is divided into a number of subspaces by the sphere set $S$. The spheres in $F_S$ overlap each other representing the compact fluid domain.
	
\begin{figure} \begin{center}
		\includegraphics[scale=0.8]{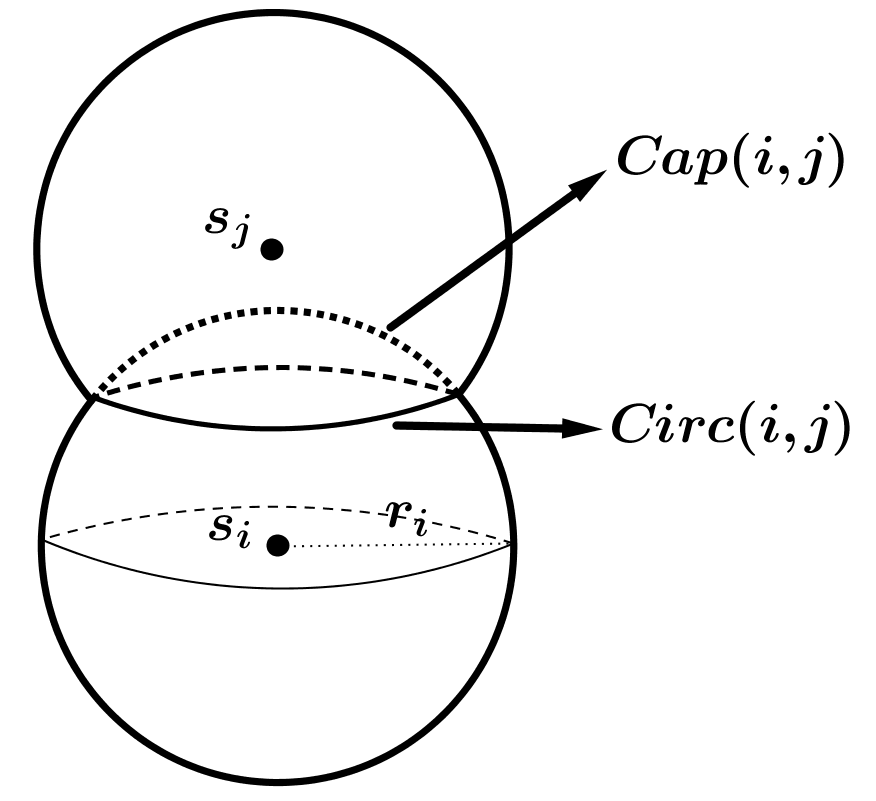}
		\caption{Illustration of terminologies}
		\label{fig:sphereCap}
	\end{center} \end{figure}

	Consider two overlapping spheres $s_i$ and $s_j$. A \emph{spherical cap} $Cap(i, j)$ of $s_j$ on $s_i$ is a spherical segment of $s_i$'s surface where every point on it is covered by $s_j$. Every $Cap(i, j)$ has to be bounded by an intersection circle $Circ(i, j)$ on the surface of $s_i$ (see figure \ref{fig:sphereCap}). The $Circ(i, j)$ is said to be \emph{covered} if all points in it are covered. 
	
	Now we consider some special cases. The radius of all spheres in $S$ is the same, so the situation that a sphere is located inside a sphere is impossible. To avoid two spheres or more to coincide, we improved the SPH algorithm with artificial viscousity\cite{Libersky} so that there is no interparticle penetration.
	
	\textbf{Definition 3.1.} Let sphere $s_i$ be overlapped by $n$ spheres. The sphere $s_i$ is said to be \emph{covered} if every point on sphere $s_i$ is covered by at least one of the $n$ intersecting spheres. The particle of the corresponding sphere is then said to be an \emph{interior} SPH particle.
	
	\begin{figure} \begin{center}
		\includegraphics[scale=0.7]{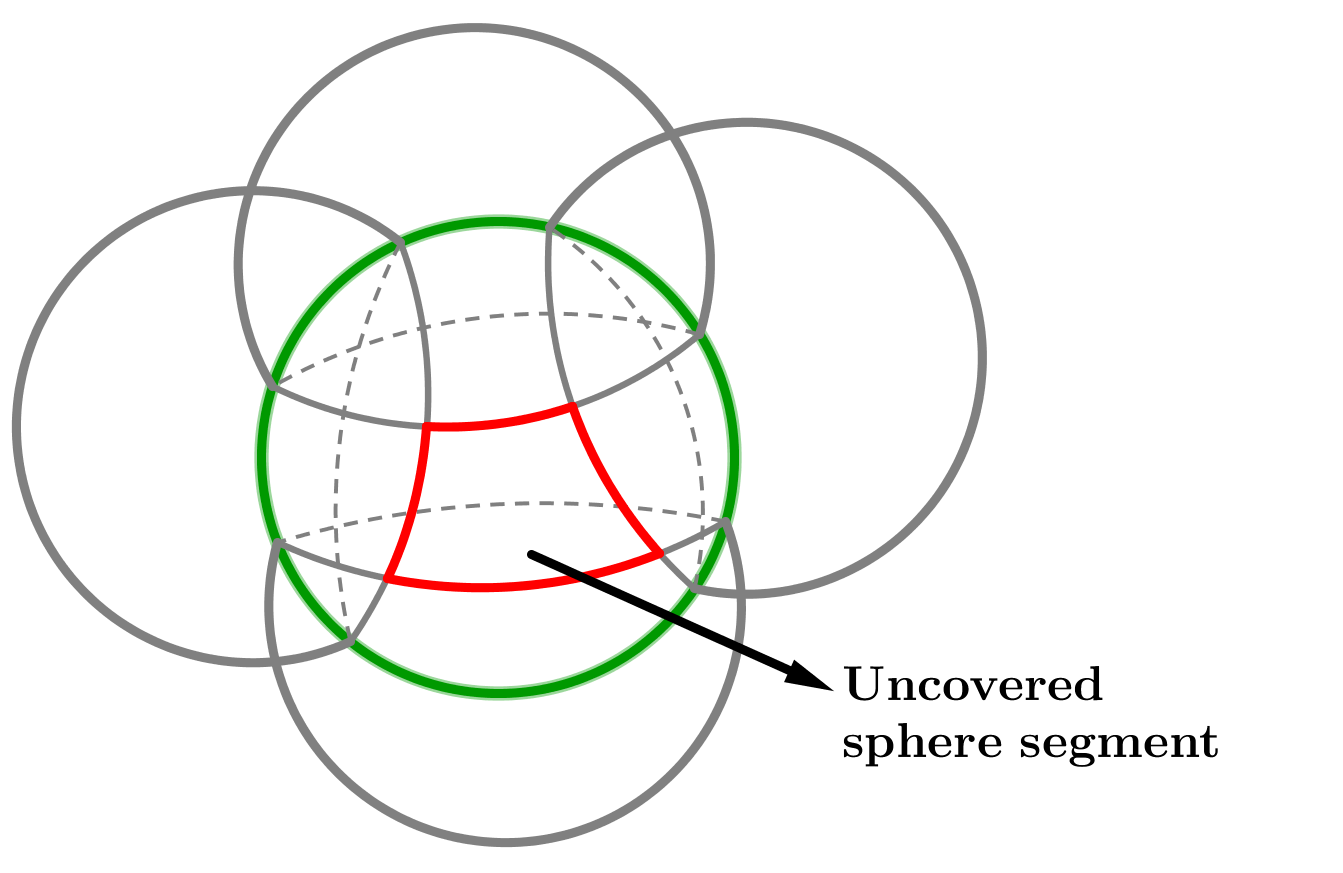}
		\caption{Illustration of a boundary particle}
		\label{fig:uncoveredSphereSegment}
	\end{center} \end{figure}	
	
	\textbf{Definition 3.2.} Let sphere $s_i$ be overlapped by $n$ spheres. The corresponding SPH particle is said to be a \emph{boundary} particle if there exists a point on the corresponding sphere which is not covered by any of the overlapping spheres (see figure \ref{fig:uncoveredSphereSegment}).
	
\begin{figure} \begin{center}
		\boxed{\includegraphics[scale=0.8]{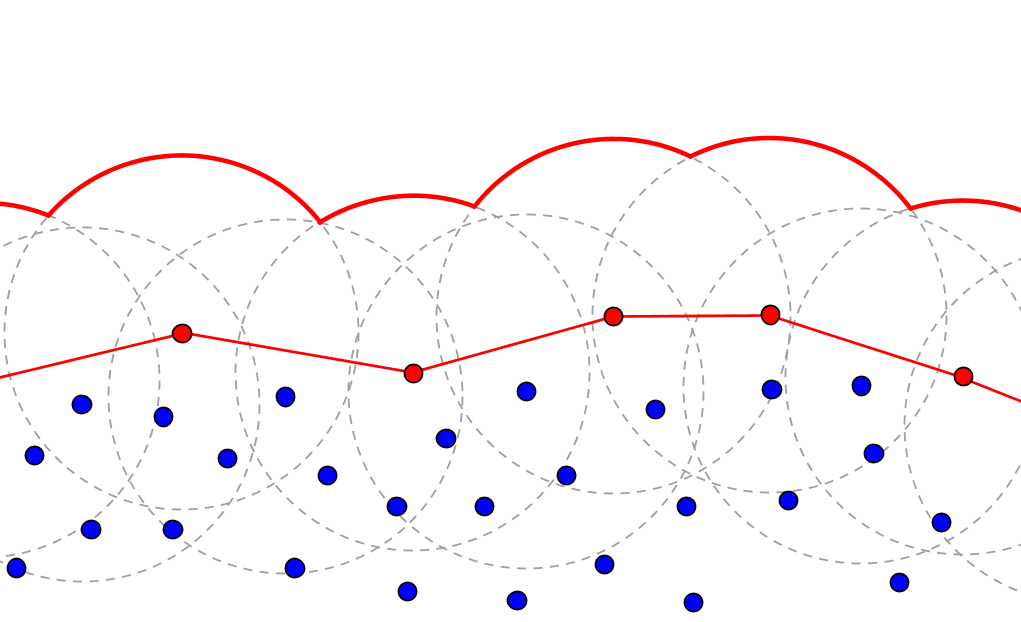}}
		\boxed{\includegraphics[scale=0.8]{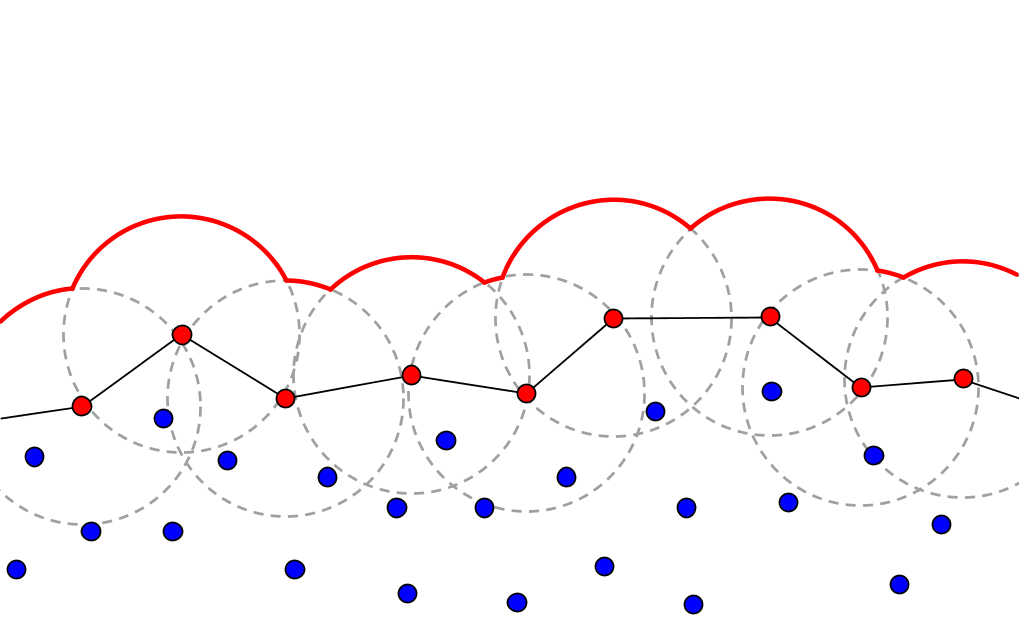}}
		\caption{Drawn in 2D, the two figures show how a smaller representative sphere radius give sharper surface profile. Uncovered sphere segments are shown in red.}
		\label{fig:boundarySmoothness}
	\end{center} \end{figure}

	The smoothness of the fluid surface depends on the value of $r_i$. If we use the spheres of smaller size, the boundary particles have greater resolution, which gives a sharper surface profile (see figure \ref{fig:boundarySmoothness}). However, a small sized sphere representation bring the risk of geometrical cavity to emerge inside the fluid. This could lead to an error of boundary detection.

	\subsection{Detection method}
	Based on the definitions above, now we explain our method to detect boundary particles.
	
\begin{figure} \begin{center}
		\includegraphics[scale=0.75]{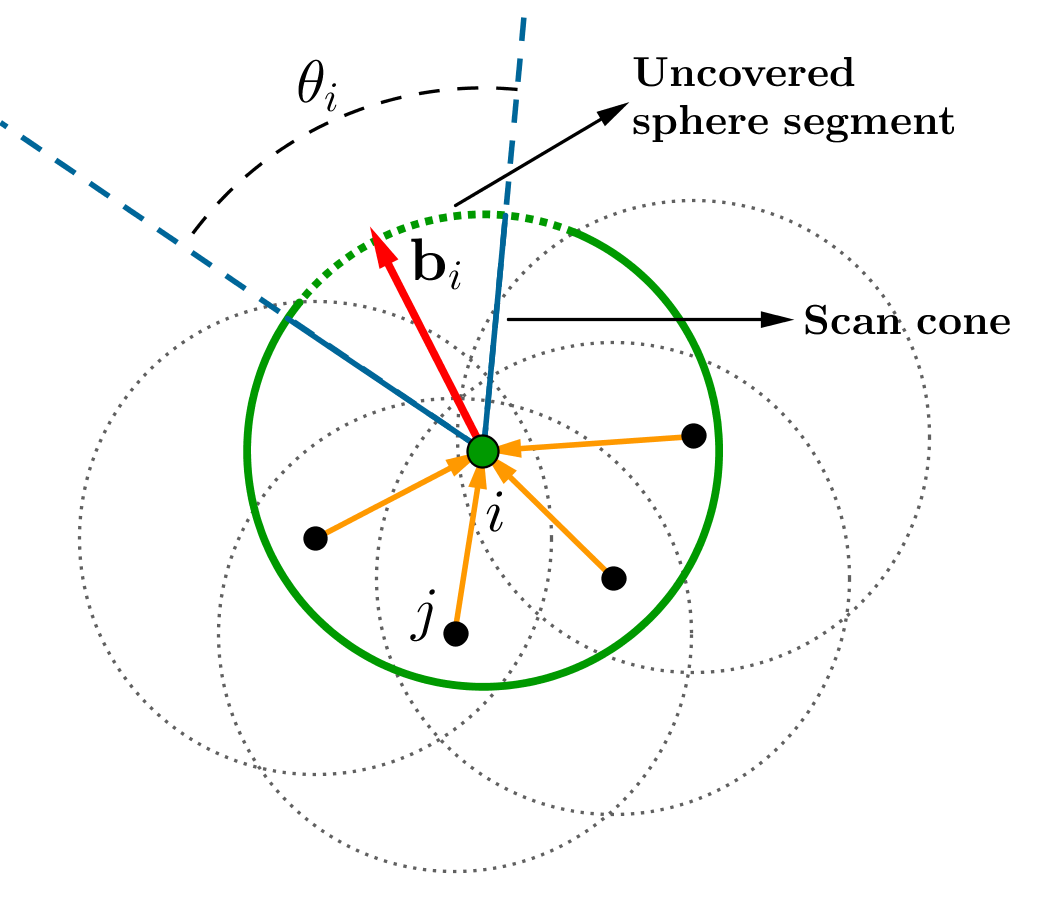}
		\includegraphics[scale=0.75]{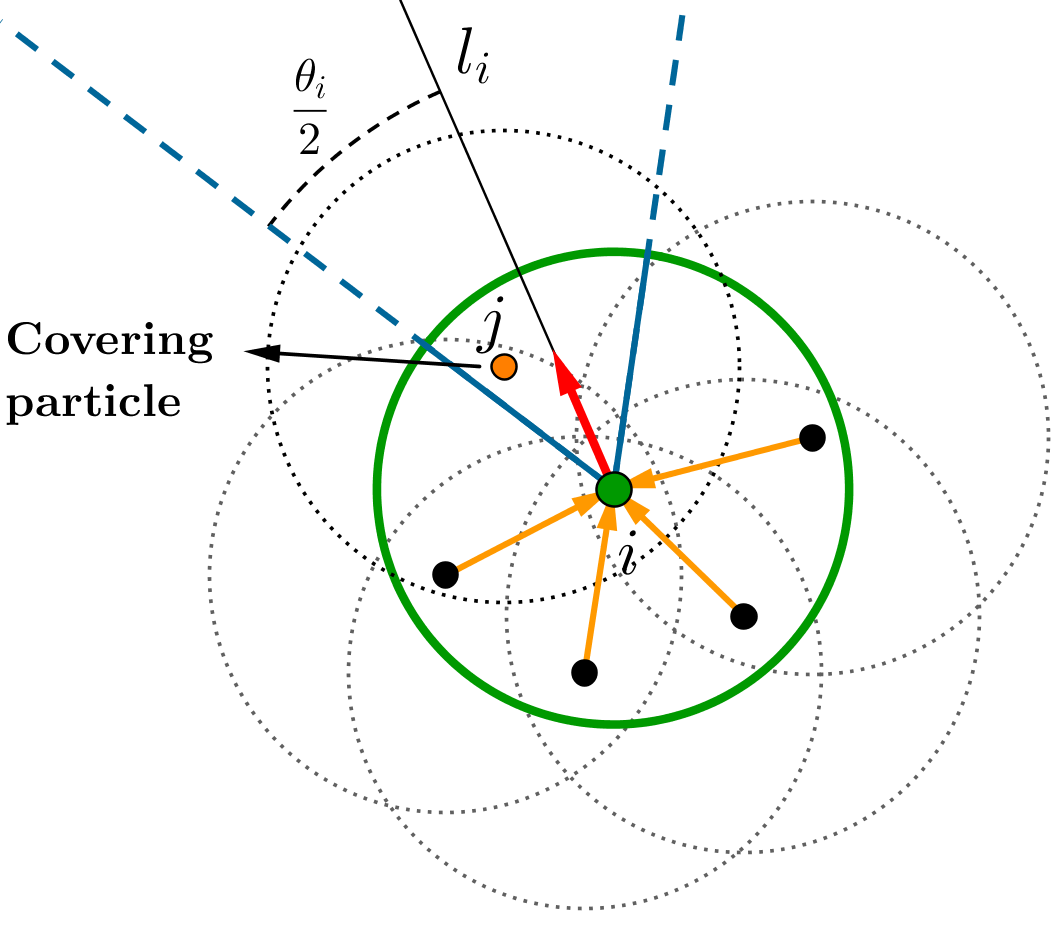}
		\caption{Illustration of cover vector and the scan cone, drawn in 2D. The cone checks if there is any other covering particle within kernel radius.}
		\label{fig:coverVectorSum}
	\end{center} \end{figure}

	Let $s_i$ be a SPH particle represented as a sphere with particle's position $\mathbf{x}_i$ as the center. Within its kernel support, $m$ neighbour particles are present to contribute for $s_i$'s SPH attributes. Let these neighbouring particle spheres overlap $s_i$. The \emph{cover vector} $\mathbf{b}_i$ is defined as (see figure \ref{fig:coverVectorSum}):
	\begin{equation}\label{eq:coverVector}
	\mathbf{b}_i = \sum\limits_{j=0}^{n} \frac{\mathbf{x}_i - \mathbf{x}_j}{|\mathbf{x}_i - \mathbf{x}_j|}
	\end{equation}
	
	Let $l_i$ be a ray that starts from $\mathbf{x}_i$ and parallel to $\mathbf{b}_i$. If $l_i$ does not intersect any spherical cap, then the corresponding sphere is not covered. Boundary particle detection can be done by checking the existence of $l_i - Cap$ intersection. 

\begin{figure} \begin{center}
		\includegraphics[scale=0.8]{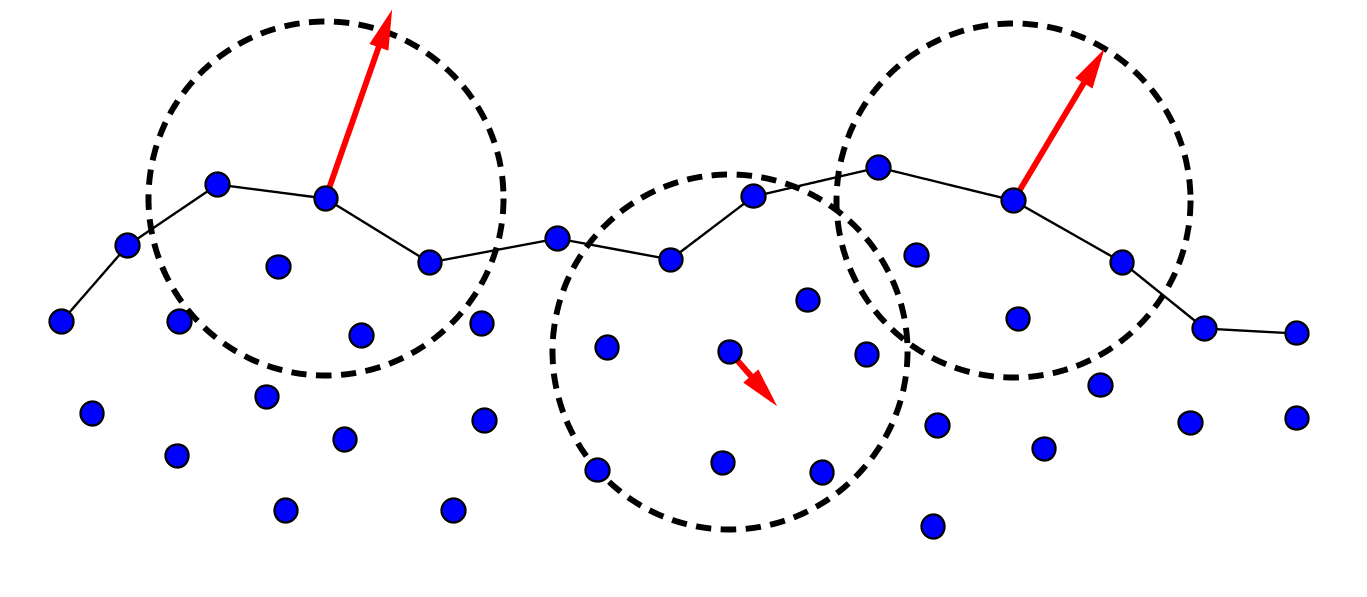}
		\caption{The cover vector on each particle differs according to the distribution of neighbouring particles.}
		\label{fig:coverVectorBehavior}
	\end{center} \end{figure}

	Now we consider the case for particle located at or near the boundary. Because of the non-uniform property of the particles (see figure \ref{fig:coverVectorBehavior}), the direction of $\mathbf{b}_i$ only serves as a rough estimation of the normal of the surface. We used a \emph{scan cone} (see figure \ref{fig:coverVectorSum}) which checks if there is any covering particle near $l_i$. The cone angle $\theta_i$ serves as threshold angle to any particle that lies near $l_i$, which detects the existence of a spherical cap intersection with $l_i$.	

	The value of $\theta_i$  depends on the radius of representative sphere used and plays important role in checking the existence of $l_i - Cap$ intersection. In our case, we directly chose the value $\theta_i = \pi/3$. The scan cone calculation check for particle $i$ is given by
	\begin{equation} \label{eq:scanCone}
	\mbox{IF } \arccos\left( \frac{\mathbf{x}_j - \mathbf{x}_i}{|\mathbf{x}_j - \mathbf{x}_i|} \cdot \frac{\mathbf{b}_i}{|\mathbf{b}_i|} \right) \le \frac{\theta_i}{2} \mbox{ THEN assign boundary status to the particle $i$}
	\end{equation} 
	
	In case when the fluid particles join to form a thin jet, a plane or become solitary particle cluster the number of neighbouring particles for each becomes very small. We considered them as boundary particles by giving a threshold number
	\begin{equation}\label{eq:neighbourThreshold}
	n_{th} =
	\begin{cases}
	4  &\mbox{ for 2 dimensional case}\\
	15 &\mbox{ for 3 dimensional case}
	\end{cases}
	\end{equation}
	to the number of neighbour particles. A particle having the number of neighbour particles greater than $n_{th}$ must be tested for its boundary status, otherwise it is a boundary particle.
	
	\subsection{Boundary detection algorithm}
	Our algorithm is embedded within the SPH solver algorithm that has been mentioned in section 2 above. The summation to calculate the cover vector can be done simultaneously with any SPH solver step that involves summation through all neighbouring particles. The simple surface detection algorithm written below uses some steps of SPH solver algorithm simultaneously:
	\begin{enumerate}
	\item At \emph{initialization}, assign all particles in the fluid domain as boundary particles
	
	\item During the step of registering particles into linked-list grid, for each particle $i$ calculate the cover vector (eq.~(\ref{eq:coverVector})) by summing \[ \frac{\mathbf{x}_i - \mathbf{x}_j}{|\mathbf{x}_i - \mathbf{x}_j|} \] for every detected neighbour particle $j$
	
	\item During the step of calculating momentum change rate for all particles (eq.~(\ref{eq:momentumDiscretized})), for each particle $i$, if the number of neighbours of $i$ is smaller than or equal to the threshold $n_{th}$ (eq.~(\ref{eq:neighbourThreshold})) then consider the particle $i$ as boundary
	
	\item During the step of calculating density change rate for all particles (eq.~(\ref{eq:continuityDiscretized})), check the existence of $l_i-Cap$ intersection by applying the scan cone calculation (eq.~(\ref{eq:scanCone})). If the intersection exists, assign the current $i$-th particle as interior.
	\end{enumerate}
	
\section{Test Cases}
We have applied the detection algorithm to several cases. To visualize particles we used spheres of radius $0.3h_{sm}$, where $h_{sm}$ is the kernel support radius. Red and blue color was used to distinguish between boundary and interior particles.

	\begin{figure} \begin{center}
		\includegraphics[scale=0.35]{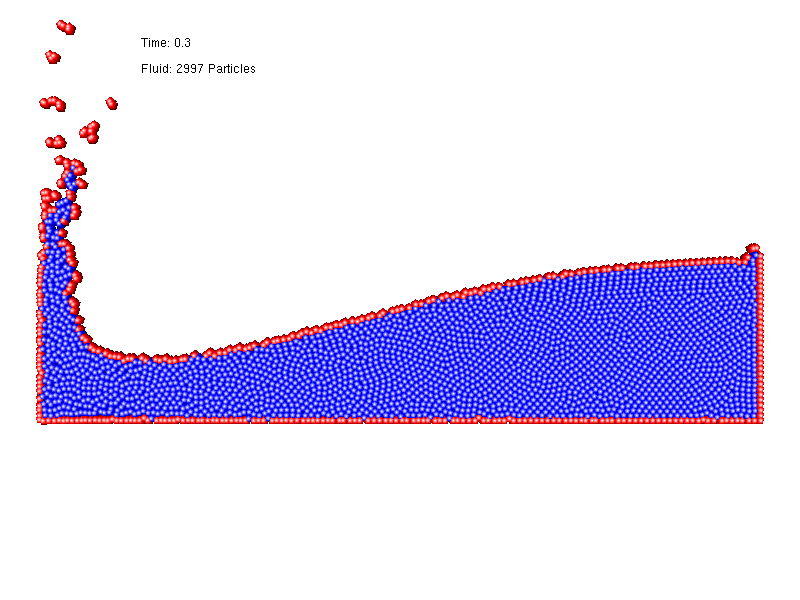}
		\includegraphics[scale=0.35]{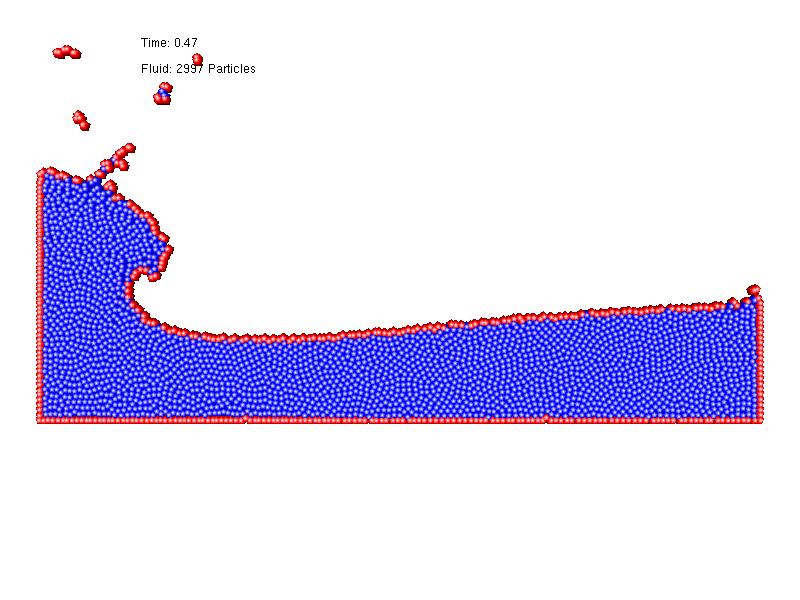}
		\caption{Two dimensional dam break}
		\label{fig:damBreak2D}
	\end{center} \end{figure}
		
	\subsection{Two-dimensional dam break}
	We implemented the algorithms to two-dimensional case so the result can be seen clearly. The simulation is a dam break in rectangular tank. The simulation consists of 2997 SPH particles with the initial shape as a rectangular block of fluid at the right side of the tank. Figure \ref{fig:damBreak2D} shows the result.

	\subsection{Two-dimensional standing wave}
\begin{figure} \begin{center}
		\includegraphics[scale=0.3]{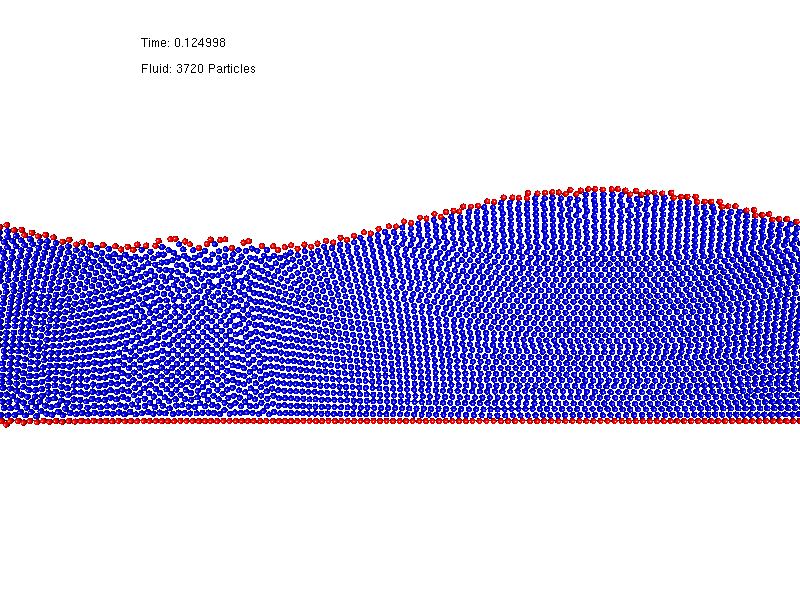}
		\includegraphics[scale=0.3]{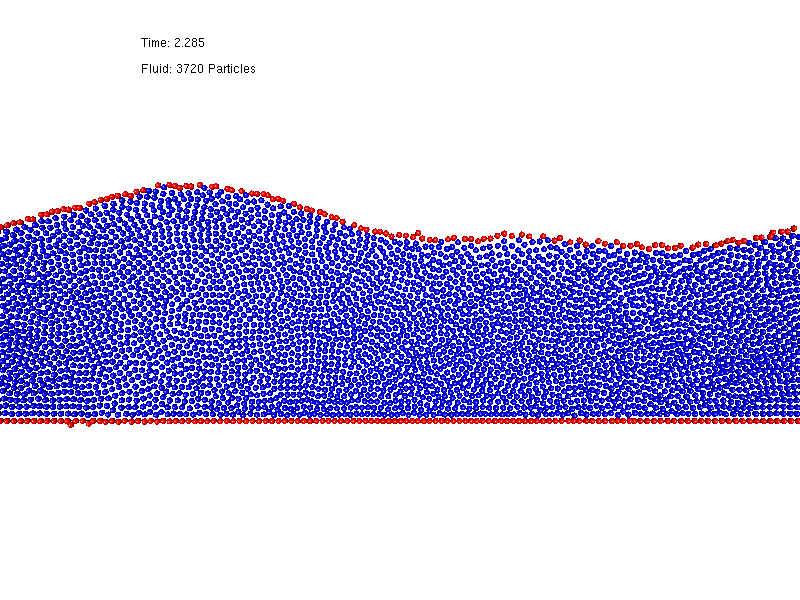}
		\caption{Two dimensional standing wave}
		\label{fig:standingWave2D}
	\end{center} \end{figure}

	The simulation is a standing wave in a rectangular tank with periodic boundary condition at the sides. The simulation consists of 3720 SPH particles with the initial condition using a sine function. Figure \ref{fig:standingWave2D} shows the result.
	
	\begin{figure} \begin{center}
		\includegraphics[scale=0.7]{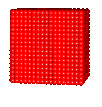}
		\includegraphics[scale=0.65]{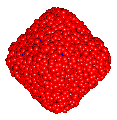}
		\includegraphics[scale=0.65]{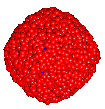}
		\includegraphics[scale=0.65]{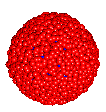}		
		\caption{Three dimensional droplet motion.}
		\label{fig:droplet3D}
	\end{center} \end{figure}

	\begin{figure} \begin{center}
		\includegraphics[scale=0.3]{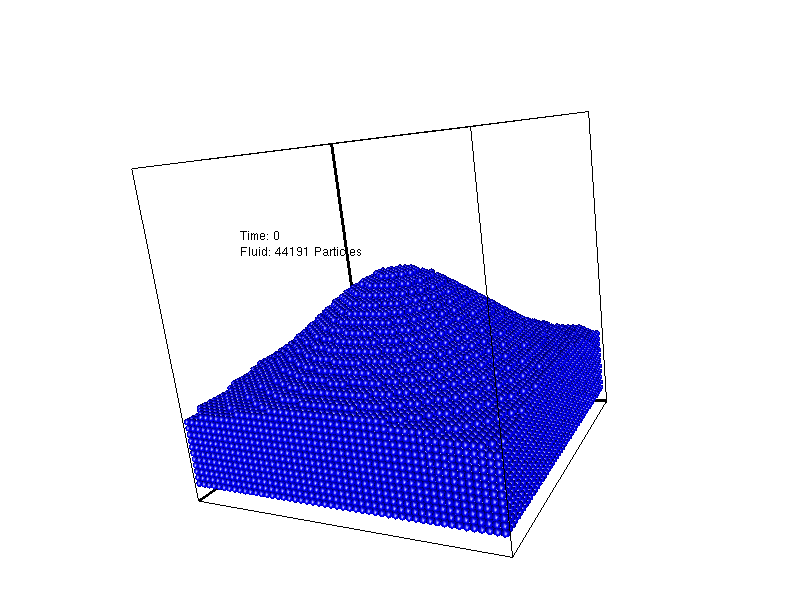}
		\includegraphics[scale=0.3]{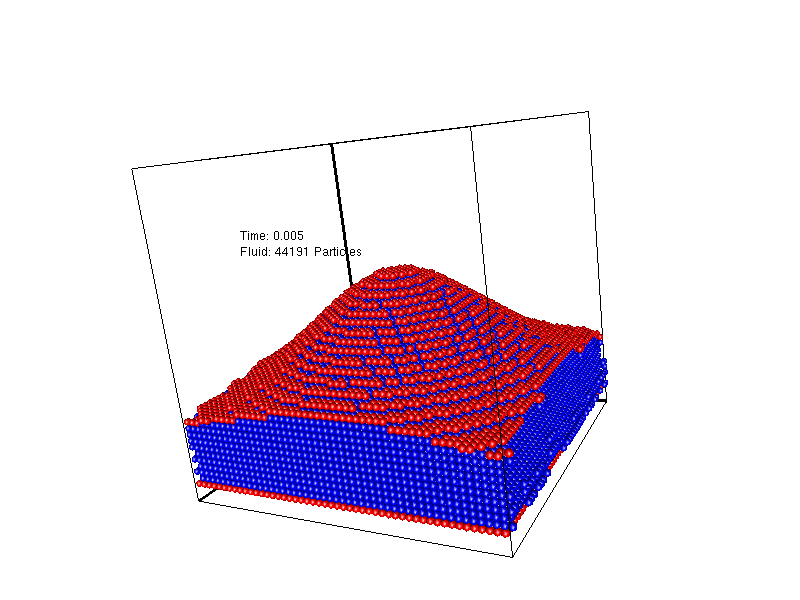}
		\includegraphics[scale=0.3]{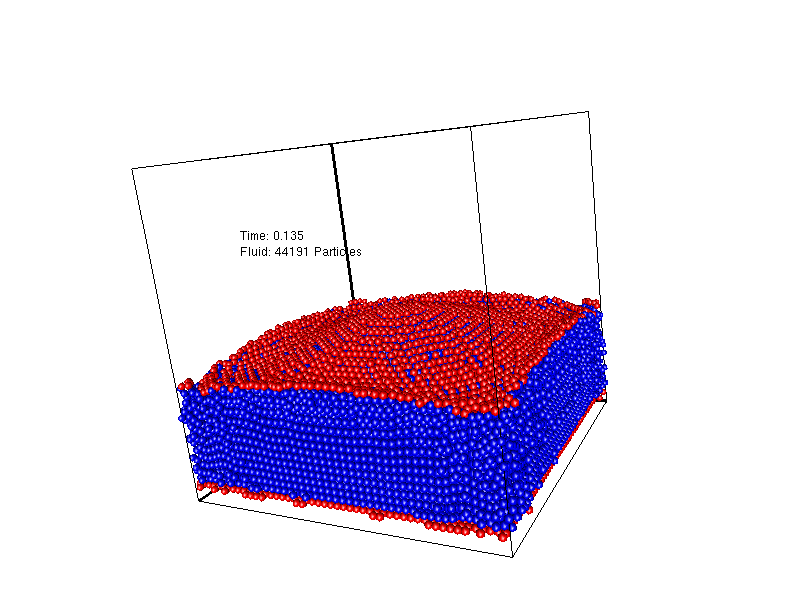}
		\includegraphics[scale=0.3]{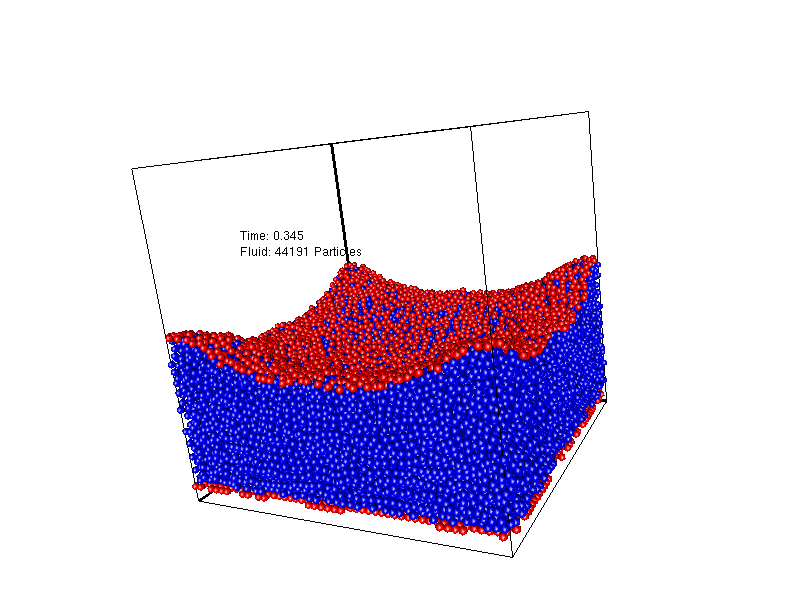}
		\caption{Three-dimensional standing wave. The upper left figure shows the system before boundary detection.}
		\label{fig:standingWave3D}
	\end{center} \end{figure}
		
	\subsection{Three dimensional droplet motion}
	This simulation applied surface tension model  used by Zhang\cite{Zhang} to the boundary particles. The droplet consists of 3375 fluid particles. The surface tension depends on the curvature of the surface shape that calculated by applying Moving Least Square fit to the boundary particles as interpolation points. Figure \ref{fig:droplet3D} shows how surface tension deforms the initial cube shape into a sphere.
	
	\subsection{Three-dimensional standing wave}
	This case is a 3-dimensional standing wave with periodic boundary condition at the sides of a cube tank. We used 44191 particles. The initialization of the dome-shaped surface is given by a gaussian function. Figure \ref{fig:standingWave3D} shows the result.

\section{Correction Method: Scan Circle}
\begin{figure} \begin{center}
		\includegraphics[scale=1.]{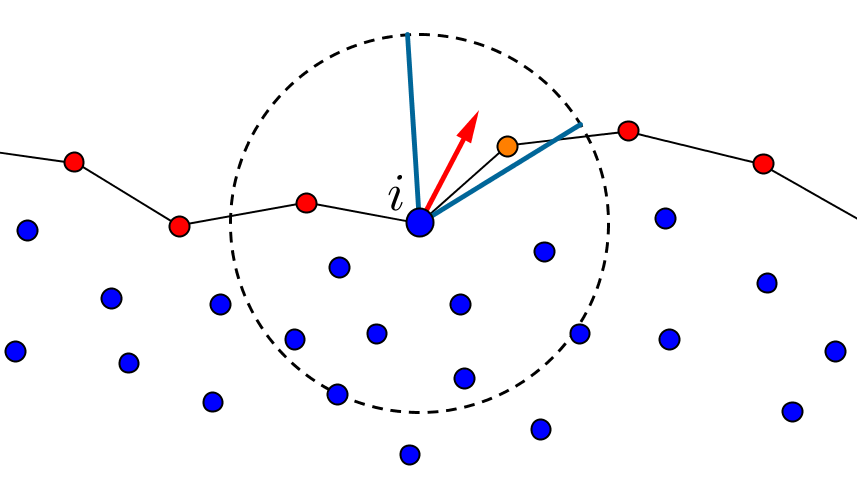}
		\caption{The scan cone of boundary particle $i$ regards another boundary particle as a covering particle. Particle $i$ is mistakenly considered as interior.}
		\label{fig:scanConeMistake}
	\end{center} \end{figure}

	The implementation of boundary detection method worked well in determining the deep interior particles, but a drawback occured when determining the boundary particles. Because of the unpredictable behavior of the neighbouring particle distribution, the cover vector of a boundary particle sometimes aims too close to another neighbour particle. The scan cone mistakenly regarded this boundary particle as interior (see figure \ref{fig:scanConeMistake}). This impairs the accuracy of this method.
		
	\begin{figure} \begin{center}
		\includegraphics[scale=0.9]{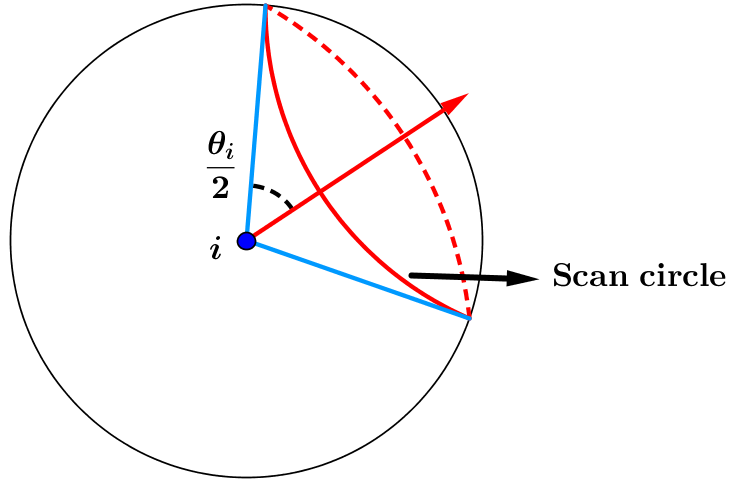}
		\caption{Scan circle}
		\label{fig:scanCircle}
	\end{center} \end{figure}
	
	To treat this problem, a further research for implementing a \emph{scan circle} is on going. Consider an intersection given by the scan cone and the sphere of a SPH particle $i$. The intersection is a circle whose center located in ray $l_i$ (see figure \ref{fig:scanCircle}). The existence of uncovered sphere segment can be tested by checking whether this circle is covered or not.

\section{Summary}
The simple boundary detection in SPH based on the implementation of cover vector and the scan cone has been successfully implemented to several cases. The algorithm can be embedded in the SPH solver algorithm easily. However, a drawback exists in the boundary detection accuracy because of random behavior of particle distribution. A research on correction method using a scan circle coverage is on going.

\end{document}